# Entanglement of a quantum optical elliptic vortex

Abir Bandyopadhyay[1, 2*], Shashi Prabhakar[1], Ravindra Pratap Singh[1†]
[1]Quantum Optics & Quantum Information Group, Theoretical Physics Division,
Physical Research Laboratory, Navrangpura, Ahmedabad – 380009, India
[2]Hooghly Engineering and Technology College, Hooghly 712103, India

**Abstract**

We calculate the entanglement of a generalized elliptical vortex formed by quantized radiation field, using Wigner quasiprobability distribution function for such states. We find a critical squeezing parameter above which the entanglement is less for higher vorticity, which is counter intuitive.

## 1. Introduction

Entanglement is fundamental in determining the usefulness of a quantum state in quantum information protocols. Entangled states play a central role in quantum key distribution, superdense coding, quantum teleportation, and quantum error correction [1], which cannot be realized classically. While a major part of the effort in quantum information theory has been in the context of systems with finite number of Hilbert space dimensions, more specifically the qubits, recently there has been much interest in the canonical continuous cases [2]. However, Gaussian states, which are part of continuous variable case, are much better studied and methods have been devised to measure the entanglement for such states [3, 4]. The measure of entanglement for a Gaussian state is best characterized by the logarithmic negativity [5], a quantity evaluated in terms of the symplectic eigenvalues of the covariance matrix [4].

Optical vortices or phase singularities in the field of light, described by a non-separable two-dimensional field, have drawn a great deal of attention in the last two decades [6]. Because of their specific spatial structure and associated orbital angular momentum (OAM), they find a variety of applications in the field of optical manipulation [7], optical communication [8], quantum information and computation [9]. For a vortex, the OAM follows the same numbers as vorticity or topological charge of the vortex i.e. an optical vortex of topological charge *m*, carries an OAM of *mℏ* per photon. Although, the OAM is quantized in case of an optical vortex, most of its studies as well as applications belong to the classical domain. It is difficult to find literature that takes into account of vortex formed by quantized radiation field. In a rare piece of work, Agarwal *et al.* have used two mode squeezed vacuum to generate circularly symmetric quantum vortex [10]. It is shown that a two mode state under the linear transformation belonging to the SU(2) group may lead to a vortex state under special conditions [11]. At this point, we must make a distinction between the optical vortices of [6] and the vortex states discussed in [10, 11] as well as considered in the present article. The optical vortices considered in [6] exist in the 2D real space of the classical wave field, while the vortex states considered in [10, 11] and in this letter are quantum mechanical wave packets formed by two quantized radiation modes and exist in the abstract space of the modes.

Recently we have generated quantum optical elliptic vortex (QEV) by coupling squeezed coherent states of two modes with beam splitter (BS) or a dual channel directional coupler (DCDC) [12]. Both the components, BS and DCDC, find practical applications in optical coherence tomography [13]. The QEV states studied by the authors can also be produced with experimental techniques which implement creation and annihilation operators [14]. To study the properties of quantum states a number of (quasi)probability distributions have been defined [15]. However, among all the (quasi)probability distributions, the Wigner function stands out, as it is real, nonsingular, yields correct quantum-mechanical operator averages in terms of phase-space integrals, and possesses positive definite marginal distributions [16]. The Wigner distribution function has come to play an ever increasing role in the description of both coherent and partially coherent beams and their passage through first order optical systems [17]. Once the Wigner distribution is known, the other properties of the system can be calculated from it. Keeping this in mind we calculated the Wigner function of the QEV [12]. We observed quantum interferences due to coupling between the two modes. In this work we discuss how this coupling could be used to generate controlled entanglement for application to quantum computation and quantum information. We show this by quantifying the entanglement in terms of the logarithmic negativity.

The paper is organized as follows. In section II we briefly describe the quantum optical elliptic vortex (QEV) and the Wigner function of it. In section III we discuss the definition of entanglement for Gaussian input states. We find the entanglement for QEV for a certain choice of parameters and discuss the result. We find a critical value of the squeezing parameter, above which higher vorticity produces less entanglement. Finally we conclude our results in section IV.

## 2. Wigner distribution of quantum elliptical vortex (QEV)

*(a) Generation of displaced quantum optical elliptical vortex (DQEV), from two squeezed coherent states, with the help of beam splitter (BS) or a dual channel directional coupler (DCDC)*

It is possible to couple two beams by using a beam splitter (BS) or a dual channel directional coupler (DCDC) [11, 12, 18, 19]. The BS is used by many authors as an entangler [18]. For two-mode states characterized by the annihilation operators $a_1$ and $a_2$, a coupling transformations can be generated by evolution under a Hamiltonian of the form $H = g(a_1^\dagger a_2 \, e^{i\varphi} + \text{h.c})$ [11]. Agarwal and Banerjee [11] constructed circular vortex state using the above mentioned Hamiltonian, described for BS/DCDC, and studied the properties of its entropy [11], while Kim et al. examined the question of the generation of entangled states by a beam splitter using Fock states as input fields [19]. Considering output operators $a_i^\dagger(out)$ are generated by the unitary transform $\mathcal{U}^\dagger a_i^\dagger(in)\mathcal{U}$, $(i = 1, 2)$, for the above mentioned Hamiltonian

$$\begin{bmatrix} a_1^\dagger(out) \\ a_2^\dagger(out) \end{bmatrix} = \begin{bmatrix} A_1 & A_2 \\ A_2 & A_1^* \end{bmatrix} \begin{bmatrix} a_1^\dagger(in) \\ a_2^\dagger(in) \end{bmatrix}, \quad (1)$$

where $\mathcal{U} = e^{-iH}$. $A_i$ denote transmitivity and reflectivity of the BS respectively and satisfy the relations $|A_1|^2 + |A_2|^2 = 1$, and $A_1^* A_2 + A_2^* A_1 = 0$. Mixing of equal amount ($A_1 = A_2$) generates a circular vortex which has been dealt quantum mechanically elsewhere [10, 11]. In the cases where the coefficients of mixing ($A_i$) are not equal, they produce an elliptical vortex. As the asymmetry becomes larger and larger, more "which path" information is available, and the quantum interference effect is correspondingly diminished. Somewhat surprisingly, this reduced interference has been found to be extremely useful in a number of quantum information processing applications in linear optics, such as, quantum computing gates [20] and quantum cloning machines [21].

The quantum mechanical description of displaced elliptic vortex in elliptic beam is given in our previous work [12]. We consider two separate squeezed coherent (displaced vacuum) modes as our input states and couple them through a BS or DCDC. At this point, we change our label for the modes from (1, 2) to (x, y). If we look at any of the output states after $m$ times the operation is performed, it is given by [12],

$$|\Psi_{qev}^D\rangle = \mathcal{N} \left[\eta_x a_x^\dagger \pm i \eta_y a_y^\dagger\right]^m S_x(\zeta_x) S_y(\zeta_y) D_x(\alpha_x) D_y(\alpha_y) |0,0\rangle, \qquad (2)$$

where $\mathcal{N}$ is the normalization constant. $S_i(\zeta_i) = exp(\zeta_i^* a_i^{\dagger 2} - \zeta_i a_i^2)$ and $D_i(\alpha_i) = exp(\alpha_i^* a_i^\dagger - \alpha_i a_i)$ are the usual squeezing and displacement operators corresponding to $x$ and $y$ directions (the index $i = x, y$). We call these states displaced quantum optical elliptical vortex (DQEV). The term in square bracket, generated by BS/DCDC, is responsible for the elliptical vortex. If we put $\eta_x = \eta_y = 1$, $\zeta_x = \zeta_y = \zeta$ (real), it reduces to the displaced circular vortex state in a displaced circular Gaussian beam (DCCV): $|\Psi_{ccv}^D\rangle$. $|\Psi_{ccv}\rangle$, a circular vortex in circular beam, is discussed in detail in [10] using $Q$ function. For the case $\zeta_x \neq \zeta_y$, the beam profile becomes elliptical, whereas $\eta_x \neq \eta_y$ refers to elliptical vortex. The parameters in the generator of the vortex term $\eta_i$ are trivially connected to the reflection and transmission of the BS, or the coupling ratios for DCDC, as described in Eq. (1). Another way to look at these states, described by Eqn. (2), is "photon added/subtracted" two squeezed vacuum states (PA/SSV) [14].

*(b) Computation of Wigner function of the QEV*

Following the mathematical treatments of [10], with the choice of the parameters, $\eta_i = 1/(\sqrt{2}\,\sigma_i)$, for $i = x, y$, one can calculate the normalized spatial distribution of displaced QEV state as [12]

$$\Psi_{qev}^D(x,y) = \sqrt{\frac{2^{(m-2)}}{\sigma_x \sigma_y \Gamma(m+\frac{1}{2})\sqrt{\pi}}} \left[\frac{(x-x_0)}{\sqrt{2}\sigma_x} \pm i \frac{(y-y_0)}{\sqrt{2}\sigma_y}\right]^m exp\left[-\frac{1}{2}\left\{\left(\frac{x-x_0}{\sigma_x}\right)^2 + \left(\frac{y-y_0}{\sigma_y}\right)^2\right\}\right], \qquad (3)$$

where, $\sigma_i = exp(2\zeta_i)$. It is centered at a point $(x_0, y_0)$, where $x_0 = \mathfrak{Re}\,(\alpha_x)$ and $y_0 = \mathfrak{Re}\,(\alpha_y)$. The distribution $|\Psi_{eev}^D(x,y)|^2$ is shown to have elliptic vortex structure, with zero intensity at $(x_0, y_0)$ i.e. the point of displacement of the vacuum [12]. Inverting the ratio $\frac{\sigma_x}{\sigma_y}$ rotates the ellipse by $\pi/2$.

Now, we change our variables to shifted (displaced) and scaled ones: $X_1 = \frac{x-x_0}{\sigma_x}$, $Y_1 = \frac{y-y_0}{\sigma_y}$, $P_{x_1} = \frac{\sigma_x}{\sqrt{2}}(p_x - p_{x_0})$, $P_{y_1} = \frac{\sigma_y}{\sqrt{2}}(p_y - p_{y_0})$, $X_2 = \frac{\sigma_y}{2\sigma_x}(x-x_0)$, $Y_2 = \frac{\sigma_x}{2\sigma_y}(y-y_0)$, $P_{x_2} = \frac{\sigma_y^3}{\sqrt{2}}(p_x - p_{x_0})$, $P_{y_2} = \frac{\sigma_x^3}{\sqrt{2}}(p_y - p_{y_0})$, with $p_{i_0} = \Im(\alpha_i)$ Following the treatment in [22], it allows us to calculate the four dimensional Wigner function for the state $|\Psi_{eev}^D\rangle$ in a compact fashion as,

$$W(x, y, p_x, p_y) = K\, exp\left[-\left(X_1^2 + Y_1^2 + P_{x_1}^2 + P_{y_1}^2\right)\right] L_m^{-1/2}\left[\frac{\left(P_{x_2} + P_{y_2} - X_2 - Y_2\right)^2}{\sigma_x^2 + \sigma_y^2}\right] \quad (4)$$

where, $L_m^{-1/2}$ is associated Laguerre polynomial (ALP), and, $K = \frac{2^{(m-4)}\, m!}{\pi\sqrt{\pi}\Gamma(m+\frac{1}{2})}\left[-2(\sigma_x^2 + \sigma_y^2)\right]^m$. We point that the effect of $\mathbf{D}_i(\alpha_i)$ is nothing but producing a displacement of the center of the beam as well as the vortex $(x_0, y_0)$. So we drop this term and call these states quantum elliptical vortex (QEV). Note that in the Eq. (4), the changed variables in the Gaussian term are different from the changed variables in the argument of the ALP term. In case of circular vortex the hole and vortex terms factor out as a product $r^{2m}L_m^0$, along with the Gaussian term. In the present case, the usual Gaussian term is factored out nicely, but the hole term ($r^{2m}$) is not separated out from the Laguerre term. We notice that it is embedded in the ALP term. Here, one can be reminded of the Rodrigues' formula for the ALP [23],

$$L_m^\alpha = \frac{(-)^m}{m!} e^{x^2} x^{-2\alpha} \frac{d^m}{dx^m}\left[e^{-x^2} x^{2(m+\alpha)}\right]. \quad (5)$$

The Eq. (5) ensures that the elliptical vortex may be expressed as a combination of circular vortices from 0 to $m$.

### 3. Entanglement for the QEV states

In this section we study entanglement of the vortex states with change in one of the squeezing parameters. The measure of entanglement has been done in terms of logarithmic negativity, which is well defined for the Gaussian states. It is well known that a two mode squeezed state or two squeezed mode state can be completely characterized by its first and second statistical moments given the covariance matrix $\Sigma$. The squeezed vacuum state also falls under the class of Gaussian states. Now, we focus on our method for generating the vortex state by the propagation of light through a BS or coupled waveguides, the unitary operations, which currently are used in quantum architectures and quantum random walks. Here, we would like to draw attention to the lemma: *"If U is a unitary map corresponding to a symplectic transformation in the phase space, i.e. if U = exp{−iH} with Hermitian H and at most bilinear in the field operators, then δA[UρU†] = δA[ρ]"* [24]. The proof of the lemma ensures that single-mode displacement and squeezing operations, as well as two-mode evolutions as those induced by a beam splitter or a parametric amplifier, *do not* change the Gaussian character of a quantum state. As the generalized vortex states, we considered, are generated only by such operations, it qualifies to be Gaussian. Note that since the first statistical moments can be arbitrarily adjusted by local unitary operations, it does not affect any property related to entanglement or mixedness and thus the behavior of the covariance matrix $\Sigma$ is all important for the study of the entanglement. Therefore, the logarithmic negativity $E_N$ [5], a quantity evaluated in terms of the symplectic eigenvalues of the covariance matrix $\Sigma$ [3, 4], and measure of the entanglement for a Gaussian state, can be applied to measure the entanglement for the QEV states. The elements of the covariance matrix $\Sigma$ are given, in terms of conjugate observables, in the symmetrized form,

$$\Sigma = \begin{bmatrix} \alpha & \mu \\ \mu^T & \beta \end{bmatrix}, \qquad (6)$$

with $\alpha = \begin{bmatrix} \langle x^2 \rangle & \langle \frac{xp_x+p_xx}{2} \rangle \\ \langle \frac{xp_x+p_xx}{2} \rangle & \langle p_x^2 \rangle \end{bmatrix}, \beta = \begin{bmatrix} \langle y^2 \rangle & \langle \frac{yp_y+p_yy}{2} \rangle \\ \langle \frac{yp_y+p_yy}{2} \rangle & \langle p_y^2 \rangle \end{bmatrix}, \mu = \begin{bmatrix} \langle \frac{xy+yx}{2} \rangle & \langle \frac{xp_y+p_yx}{2} \rangle \\ \langle \frac{yp_x+p_xy}{2} \rangle & \langle \frac{p_xp_y+p_yp_x}{2} \rangle \end{bmatrix}.$

The structure of $\Sigma$ ensures that it is the transpose of itself ($\Sigma^T = \Sigma$). The symmetric operator averages in the matrix elements of $\Sigma$ are calculated from the Wigner function using the relation $\langle \hat{O} \rangle = \iint_{-\infty}^{\infty} dx\, dp_x \iint_{-\infty}^{\infty} dy\, dp_y\, \hat{O}\, W(x,y,p_x,p_y)$. The condition for entanglement of a Gaussian state is derived from the Peres-Horodecki positive partial transpose (PPT) criterion [4], according to which the smallest symplectic eigenvalue $\nu_<$ of the transpose of matrix $\Sigma$ should satisfy,

$$\nu_< < 1/2, \qquad (7)$$

where, $\nu_< = \min[\nu_+,\ \nu_-]$. In this definition eigenvalues ($\nu_+,\ \nu_-$) are given by

$$\nu_\pm = \sqrt{\frac{\Delta(\Sigma) \pm \sqrt{\Delta(\Sigma)^2 - 4\mathrm{Det}\Sigma}}{2}} \qquad (8)$$

where, $\Delta(\Sigma) = \mathrm{Det}(\alpha) + \mathrm{Det}(\beta) - 2\mathrm{Det}(\mu)$. Thus according to the condition in Eqn. (7), when $\nu_< \geq 1/2$, Gaussian states become separable. The corresponding quantification of entanglement is given by the logarithmic negativity $E_N$ [5] defined as,

$$E_N = \max[0, -\ln(2\nu_<)] \qquad (9)$$

At this point we note an interesting observation from the Figs (5) and (6) in Ref [25], which is not mentioned by the authors of the said reference. We note that in both the plots the entanglement show similar periodicity over mixing ratio, but lagging/leading with a phase factor of $\pi/2$ [25], as outcome of BS/DCDC coupling [11, 12, 18, 19]. The separable state acquires entanglement due to the action of BS/DCDC. Especially in DCDC the entangled state again becomes separable after certain distance, and the process repeats periodically. It is interesting to note similar oscillation from one to the other type has also been reported elsewhere [11]. It means that through DCDC a two (separate) squeezed state mode (TSSM) is converted to a two mode squeezed state (TMSS) and again to TSSM periodically. Our states, described by the Wigner function in Eqn. (5), describe both the extreme cases, along with the generalized states, in between.

Remembering the fact that the effect of $D_i(\alpha_i)$ is nothing but shifting the center of the beam as well as the vortex, we choose $x_0 = y_0 = p_{x_0} = p_{y_0} = 0$, in the Wigner function, without loss of any new information, and compute the dependence of entanglement on squeezing parameter. The states in between may also be described by the QEV state, expressed by Eqns. (2-4). We report a *critical* squeezing parameter, above which, higher AM means lower entanglement. We have computed the entanglement for a choice of parameters $\sigma_y = \sqrt{5\sigma_x}$. In terms of $\zeta_i$, the relationship is linear: $\zeta_y = \frac{\ln 5}{4} + \frac{\zeta_x}{2}$. We have plotted the entanglement, $E_N$, in Fig. (1), for $m = 0$ to 5 i.e. for different orders of the vortex, as a function of $\sigma_x$. First of all, we analyze our observation for $m = 0$. For this state, we observe entanglement, which is counter intuitive. $m = 0$ means that no vortex is formed, thus there should be no entanglement, as it is TSSM. The reason for this apparent contradiction is explained below. It is definitely true that if the two squeezing parameters are random, then the state would be separable. However, we suspect that due to our (or, logically speaking, any) choice of a specific relation between the squeezing parameters, some sort of entanglement is generated. Thus we argue that the constant entanglement, generated in our computation, is due to the non-random choice of squeezing parameters. The observation of the constant value of the entanglement supports the logic that as it is generated with some fixed relationship, it remains constant. We have verified the fact that some other fixed relationship produces entanglement, with some other constant value. However, the other dependencies of the parameters will be considered in future correspondences, if found with interesting features.

Similar properties are again evident from the plot of entanglement vs. $\zeta_x$, in Fig. (2). The intercepts of the plots vary as $\ln K$, defined after Eq. (4). The important observation in both the figures is that the above a critical point, $\sigma_x = 0.002$ ($\zeta_x = -3.1073 = 3.1073\, e^{i\pi}$), the higher charge or the higher OAM corresponds to lower entanglement. The squeezing parameter $\zeta_i$ is complex in general. However, most of the studies consider only real positive values of the parameter. We report our work in the complex domain of the parameter from the expression of $\zeta_x$ mentioned above, where the negativity is realized in the phase factor of the complex squeezing parameter. A critical value of the squeezing parameter, above which the resolution of the Mach-Zehnder interferometer decreases, has been reported [26] previously also. It implies that the different domains of the generally complex squeezing parameters should be explored for the different experimental setups.

## 4. Conclusions

To conclude, we proposed the two well known mechanisms (BS and DCDC) of coupling between two squeezed and displaced modes and a recent one (PA/SSV) to generate a quantum optical elliptic vortex (QEV). We have argued that the QEV is a Gaussian state as squeezing or coupling between the two modes do not change this property. Thus the entanglement follows the Peres-Horodecki PPT criterion and the logarithmic negativity of the lowest eigenvalue of the covariance matrix. We have computed the entanglement of such quantum elliptical vortex from the four dimensional Wigner distribution function, which is used to find out covariance matrix and therefore, the logarithmic negativity. We show that by changing the squeezing parameter one can control the entanglement. We observed a critical point above which the increase in vorticity decreases the entanglement; the point corresponds to a negative real value of the squeezing parameter.

**Acknowledgements**

AB acknowledges the Associateship at PRL.

___________________________________________________________________________
*Email: abir@hetc.ac.in; †Email: rpsingh@prl.res.in

# **Figure Captions**

Fig. 1. Plot of entanglement $E_N$ (in arbitrary units) versus $\boldsymbol{\sigma_x}$, for different charge (*m*) of QEV. Note that the finite constant entanglement for *m* = 0 (explained in the text) and critical point of squeezing parameter, where the curves $\boldsymbol{m \neq 0}$ cross each other.

Fig. 2. Plot of entanglement $E_N$ (in arbitrary units) versus $\boldsymbol{\zeta_x}$, for different charge (*m*) of QEV. The exponential in Fig. (1) is clearly become linear. The intercepts varies as $\ln K$, defined after Eqn. (4).

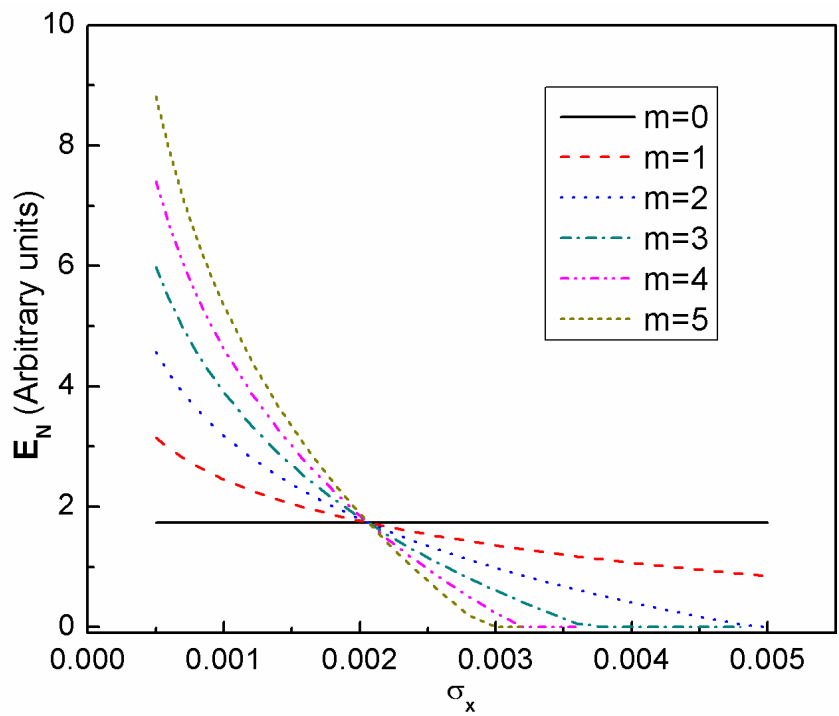

**Figure 1**

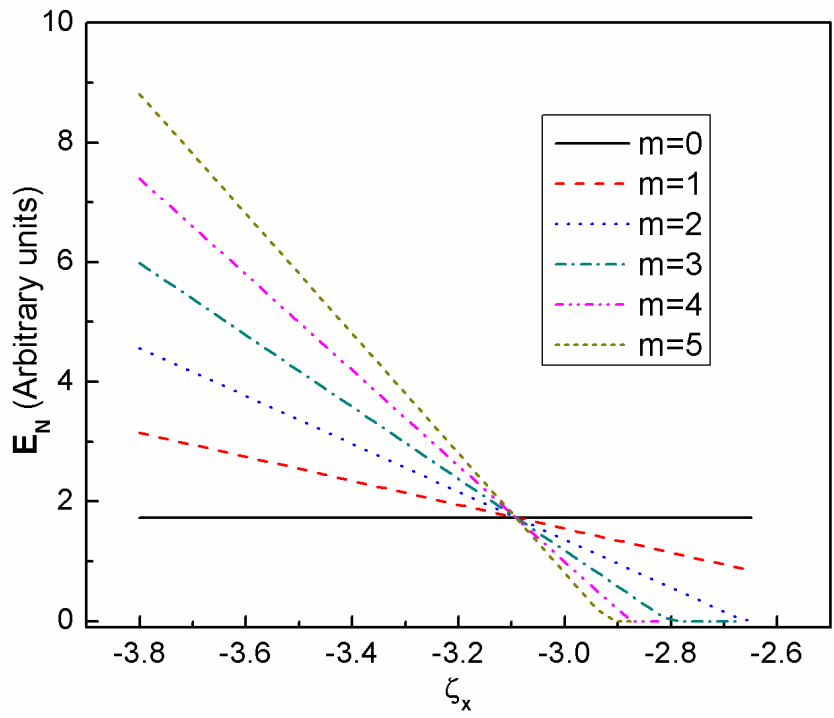

**Figure 2**